\documentstyle[11pt]{article}
\def \bA {{\bar A}}
\def \bdel {{\bar \partial}}
\def \d {{\delta}}
\def \bd {\bar{\partial}}
\def \del {{\partial}}
\def \bD {{\bar D}}
\def \bz {{\bar z}}
\def \A {{\cal A}}

\def \G {{\cal G}}
\def \H {{\cal H}}
\def \half {{\textstyle{1\over 2}}}
\def \la {{\langle}}

\def \ra {{\rangle}}
\def \Tr {{\rm Tr}}
\def \vf {{\varphi}}

\def \vx {{\vec x}}
\def \by {{\bar y}}
\def \S {{\cal S}}

\newcommand{\be}{\begin{equation}}
\newcommand{\ee}{\end{equation}}
\newcommand{\no}{\nonumber}
\begin{document}

\begin{titlepage}
\null\vspace{-62pt}

\pagestyle{empty}
\begin{center}
\rightline{}
\rightline{CCNY-HEP-01/09}

\vspace{1.0truein} {\Large\bf Yang-Mills theory in (2+1) 
dimensions: a short review}

\vspace{1in} V. P. NAIR\\
\vskip .1in {\it Physics Department\\ City College of the CUNY\\
New York, NY 10031}\\
\vskip .05in {\rm E-mail: vpn@sci.ccny.cuny.edu}\\
\vspace{1.5in}

\centerline{\large\bf Abstract}
\end{center}
The analysis of (2+1)-dimensional Yang-Mills ($YM_{2+1})$ 
theory via the use 
of gauge-invariant matrix variables is reviewed. 
The vacuum wavefunction, 
string tension,
the propagator mass for gluons, its relation to 
the magnetic mass for $YM_{3+1}$ at nonzero temperature
and the extension of our analysis to the
Yang-Mills-Chern-Simons theory are discussed.
(Talk given at the Lightcone Workshop, Trento, 2001,
to be published in Nucl. Phys. Proceedings and Supplements.)

\end{titlepage}

\pagestyle{plain}
\setcounter{page}{2}
\baselineskip =16pt

\section{INTRODUCTION}

In this talk I shall discuss a Hamiltonian approach to Yang-Mills theory in two spatial dimensions
($YM_{2+1}$), where nonperturbative calculations can be carried out to the extent that results
on mass gap and string tension can be compared with lattice simulations of the theory.
The work I shall report on was developed over the last few years in collaboration with 
D. Karabali and
Chanju Kim \cite{1,2,3}.

Before entering into the details of our work, let me say a few words about the relevance
of $YM_{2+1}$. Gauge theories (without matter) in (1+1) dimensions are rather trivial since
there are no propagating degrees of freedom, although there may be global degrees of
freedom on spaces of nontrivial topology. In (2+1) dimensions, gauge theories do have
propagating degrees of freedom and being next in the order of complexity, it is possible
that they provide a model simple enough to analyze mathematically and yet nontrivial
enough to teach us some lessons about (3+1)-dimensional
$YM$ theories. Another very important reason to study $YM_{2+1}$ is its relevance to
magnetic screening in $YM_{3+1}$ at high temperature. Gauge theories at finite temperature
have worse infrared problems than at zero temperature due to the divergent nature of the
Bose distribution for low energy modes. A dynamically generated Debye-type screening mass
will eliminate some of these, but we need a magnetic screening mass as well to have a
perturbative expansion which is well defined in the infrared.

A simple way to see that a magnetic mass can be dynamically generated
is as follows.
Consider the imaginary time
formalism, with Matsubara frequencies $\omega_n = 2\pi n T$, where $T$ is the
temperature. At high temperatures and for
modes of wavelength long compared to $1/T$, the modes with nonzero 
$\omega_n$ are
unimportant and the theory reduces to the theory of the $\omega_n =0$ mode, viz., a
three (Euclidean) dimensional Yang-Mills theory (or a (2+1)-dimensional theory in
a Wick rotated version). Yang-Mills theories in three or (2+1) dimensions are expected to have
a mass gap and this is effectively the magnetic mass of the (3+1)-dimensional
theory at high temperature \cite{lingpy}. 

In the following, it is useful to keep in mind certain
facts about $YM_{2+1}$. The
coupling constant
$e^2$ has the dimension of mass and it does not run as the four-dimensional
coupling does. The dimensionless expansion parameter of the theory is
$k/e^2$ or $e^2/k$, where $k$ is a typical momentum. Thus modes of low momenta
must be treated nonperturbatively, while modes of high momenta can be treated 
perturbatively. There is no simple dimensionless expansion parameter.
$YM_{2+1}$ is perturbatively super-renormalizable, so the ultraviolet  
singularities are well under control. 

\section{PARAMETRIZATION OF FIELDS}

Consider 
a gauge theory with group $G=SU(N)$ in the $A_0 =0$ gauge. The gauge potential 
can be written as $A_i = -i t^a A_i ^a$, $i=1,2$, where $t^a$ are hermitian  
$N \times N$-matrices which form a basis of the Lie algebra of $SU(N)$ with 
$[t^a, t^b ] = i f^{abc} t^c,~~{\rm {Tr}} (t^at^b) = {1 \over 2} \delta ^{ab}$. 
The spatial coordinates $x_1 ,x_2$ will be combined into the complex combinations 
$z=x_1 -ix_2,~{\bar z} =x_1+ix_2$ with the corresponding components for the
potential 
$A\equiv A_{z} = {1 \over 2} (A_1 +i A_2), ~~  
{\bar A}\equiv A_{\bar{z}} = 
{1 \over 2} (A_1 -i A_2) = - (A_z)^{\dagger}$. 
The starting point of our analysis is a change of variables given by
\be 
A_z = -\partial_{z} M M^{-1},~~~~~~~~~~~~~ A_{\bar{z}} = M^{\dagger -1} \partial_ 
{\bar{z}} M^{\dagger}   
\label{1}
\ee 
Here $M,~M^\dagger$ are complex matrices in general, not unitary. If they are unitary,
the potential is a pure gauge. 
The parametrization (\ref{1})
is standard in many discussions of two-dimensional gauge fields. 
A particular advantage of this parametrization is the way gauge transformations are
realized. A gauge transformation $A_i \rightarrow 
A_i^{(g)} = g^{-1} A_i g + g^{-1} \partial_i g, ~g(x)\in SU(N)$ is obtained 
by the transformation $M\rightarrow M^{(g)}=g M$. The gauge-invariant degrees of freedom
are parametrized by the hermitian matrix $H=M^\dagger M$.
Physical state wavefunctions are functions of $H$.

In making a change of variables in a Hamiltonian formalism, there are two things we
must do: 1) evaluate the volume measure (or Jacobian of the transformation) which
determines the inner product of the wavefunctions and 2) rewrite the Hamiltonian as an
operator involving the new variables. A consistency check would then be the
self-adjointness of the Hamiltonian with the given inner product. We begin with the
volume measure for the configuration space.

\section{THE INNER PRODUCT}

The measure of integration over the fields $A, \bA$ is
${d\mu ({\A})/ vol({\G}_*)}$
where
$d\mu ({\cal A})= \prod_{x,a} dA^a (x) d\bA^a (x)$ is the
Euclidean volume element on the space of gauge potentials ${\cal A}$ and  
$vol{\cal G_*}$ is the volume of gauge transformations,
viz., volume of $SU(N)$-valued functions on space.
From (\ref{1}) we see that
\begin{eqnarray}
\delta A&=& -D(\delta M M^{-1})\no\\
&=& -\left( \partial (\delta M M^{-1} )
+[A, \delta M M^{-1}]~\right)
\nonumber\\
\delta \bA &=& \bD (M^{\dag -1}\delta M^\dag ) \label{2}
\end{eqnarray}
which gives
\be
d\mu ({\cal A}) = (\det D \bD )~d\mu (M, M^\dag )
\label{3}
\ee
where $d\mu (M, M^\dag )$ is the volume for the complex matrices $M, M^\dag$, which is associated
with the metric $ds_M^2~=8 \int {\Tr}(\delta M
M^{-1}~M^{\dagger -1} \delta M^\dagger )$. This is given by the highest order differential form
$dV$ as $d\mu (M, M^\dag )= \prod_x dV(M,M^\dag )$ where
\begin{eqnarray}
dV(M,M^{\dagger}) \propto &&\!\!\!\!\!\!\!\!\!\!\! \epsilon _{a_1...a_n} 
(dM M^{-1})_{a_1}\!\!\!
\wedge ...(dMM^{-1})_{a_n}  \nonumber\\
 &&\!\!\!\!\!\!\!\!\!\!\!\!\! \times  \epsilon _{b_1...b_n} (M^{\dagger
-1} d M^{\dagger})_{b_1}
 \cdots\no\\
\label{4}
\end{eqnarray}
where $n={\rm dim} G={\rm dim} SU(N) = N^2 -1$. (There are some
constant numerical factors which are irrelevant for our discussion.)
The complex matrix $M$ can be written as $M= U \rho$, where $U$ is unitary and $\rho$ is hermitian.
This is the matrix analogue of the modulus and phase decomposition for a complex number.
Since gauge transformations act as $M\rightarrow M^{(g)}=g M$, we see that $U$ represents the gauge
degrees of freedom and $\rho$ represents the gauge-invariant degrees of freedom on $M$.
Substituting $M = U\rho$, (\ref{4}) becomes
\begin{eqnarray}
dV(M, M^{\dagger}) \propto  &&\!\!\!\!\!\!\!\!\!\! \epsilon _{a_1...a_n}
(d\rho
\rho^{-1} +
\rho^{-1} d\rho)_{a_1} \wedge ...  \nonumber\\
&&\!\!\!\!\!\!\!\!\!\!\times \epsilon _{b_1...b_n} (U^{ -1} d U)_{b_1}
\wedge\cdots  \nonumber\\
\propto &&\!\!\!\!\!\!\!\!\!\!  d\mu (U)\epsilon _{a_1...a_n}
(H^{-1}dH)_{a_1}
\wedge \cdots \no\\
\label{5}
\end{eqnarray}
Here $d\mu (U)$ is the standard group volume measure (the  Haar measure)
for $SU(N)$.
Upon taking the product over all points, $d\mu (U)$ gives the
volume of the entire gauge group, namely 
$vol(\G_*)$, and thus
\begin{eqnarray}
d\mu (M, M^{\dagger}) &&\!\!\!\!\!\!\!\!\!= \prod_{x} dV(M, M^{\dag}) 
~ vol (\G _*) \no\\
&&\!\!\!\!\!\!\!\!\!= d\mu (H)  ~vol (\G _*) \label{6}\\
d\mu (H)&&\!\!\!\!\!\!\!\!\! =
= \epsilon _{a_1...a_n}
(H^{-1}dH)_{a_1} ... (H^{-1}dH)_{a_n}\no\\ 
\end{eqnarray}
The volume element or the integration measure for the gauge-invariant configurations 
can now be written as
\begin{eqnarray}
{d\mu ({\A})\over vol({\G}_*)}
=&&\!\!\!\!\!\!\!\!\! {[dA_z dA_{\bar{z}}]\over vol({\G}_*)} \no \\
=&&\!\!\!\!\!\!\!\!\! (\det D_z D_{\bar{z}}) {d\mu  (M,
M^{\dagger})\over vol({\G}_*)} \no\\
=&&\!\!\!\!\!\!\!\!\! (\det D \bD ) d\mu (H)\label{7}
\end{eqnarray}
where we have used (\ref{6}).
The problem is thus reduced to the calculation of the determinant of the
two-dimensional operator $D\bD$. This is well known \cite{poly}. 
The simplest way to evaluate this is to define $\Gamma = \log~\det D\bD$, which gives
\be
{\delta \Gamma \over \delta \bA^a}~= -i~\Tr\left[ \bD^{-1}(x,y) 
T^a\right]_{y\rightarrow x}\label{7b}
\ee
$(T^a)_{mn}=-if^a_{mn}$ are the generators  of the Lie algebra in the adjoint
representation. The coincident-point limit of $\bD^{-1}(x,y)$ is
singular and needs regularization. With a gauge-invariant regulator,one finds
\begin{eqnarray}
\Tr &&
\!\!\!\!\!\!\!\!\!\left[ \bD^{-1}_{reg}(x,y) T^a \right]_{y\rightarrow x}\no\\
&&~~~~~~~~~= {2c_A \over
\pi}
\Tr \left[ (A -M^{\dag -1} \partial M^\dag )t^a\right]\no\\
\label{7c}
\end{eqnarray}
where $c_A \delta^{ab} = f^{amn}f^{bmn}$; it is equal to $N$ for $SU(N)$.
Using this result in (\ref{7b}) and integrating we get 
\begin{eqnarray}
(\det D \bD) &&\!\!\!\!\!\!\!\!\!= \left[ {{\det ' \del \bdel } \over \int d^2 x}
\right] ^{{\rm dim} G} ~ \exp \left[ 2c_A ~\S (H) \right]\no\\
\label{8}
\end{eqnarray}
$\S (H)$ is the 
Wess-Zumino-Witten (WZW) action for the hermitian matrix field $H$ given by \cite{witt}
\begin{eqnarray}
{\S} (H) &&\!\!\!\!\!\!\!\!\!= {1 \over {2 \pi}} \int \Tr (\partial H \bar{\partial}
H^{-1}) +{i
\over {12 \pi}} \int \epsilon ^{\mu \nu \alpha} \times\no\\
&&~~~\Tr ( H^{-1} \partial _{\mu} H  H^{-1}
\partial _{\nu}H H^{-1} \partial _{\alpha}H)\no\\
 \label{9}
\end{eqnarray}

We can now write the inner product for states $|1\ra$ and
$|2\ra$, represented by the wavefunctions $\Psi_1$ and $\Psi_2$, as \cite{GKBN}
\be
\label{inprod}
\la 1 | 2\ra = \int d\mu (H) e^{2c_A ~\S (H)}~~\Psi_1^* \Psi_2 \label{10}
\ee

\section{THE HAMILTONIAN}

The next step is the change of variables in the Hamiltonian. However, 
there is some
further simplification we can do before taking up the Hamiltonian. 
We would expect the wavefunctions to be functionals of the matrix field
$H$, but actually we can take them to be functionals of the current of the WZW model
(\ref{9}) given by
$J= (c_A/\pi ) \partial_zH~H^{-1}$. 
Notice that matrix elements calculated with (\ref{10}) are correlators of the
hermitian WZW model of level number $2c_A$. The properties of the
hermitian model of level number $k+2c_A$ can
be obtained by comparison  with the $SU(N)$-model defined by $e^{k
\S (U)},~ U(\vx)\in SU(N)$.  The hermitian analogue of the renormalized level $\kappa
= (k+c_A)$ of the
$SU(N)$-model is $-(k+c_A)$. Since the correlators involve only the  renormalized
level
$\kappa$, we see that the correlators of the  hermitian model 
(of level $(k+2c_A)$ ) can
be obtained from the correlators of  the
$SU(N)$-model (of level $k$ ) by the analytic continuation 
$\kappa \rightarrow -\kappa$. For the $SU(N)_k$-model there are the  so-called
integrable representations whose highest weights are limited  by $k$ (spin $\leq k/2$
for $SU(2)$, for example). Correlators involving  the nonintegrable representations
vanish. For the hermitian model the  corresponding statement is that the correlators
involving nonintegrable  representations are infinite.
In our case, $k=0$, and we have only one integrable representation corresponding to the
identity operator (and its current algebra descendents). Therefore, for states of
finite norm, it is sufficient to consider $J$ \cite{1,2}.

This means that we can transform the Hamiltonian ${\cal H}= T+V$
to express it in terms of
$J$ and functional derivatives with respect to $J$.
By the chain rule of differentiation
\begin{eqnarray}
T\Psi&&\!\!\!\!\!\!\!\!\! = {e^2\over 2}\int E^a_iE^a_i ~\Psi\no\\
&&\!\!\!\!\!\!\!\!\!= -{e^2\over 2}\Biggl[ \int_{x,u}{\delta  J^a ( u )\over \delta
A^c_i(x)\delta A^c_i(x)} {\delta \Psi
\over \delta J^a(u)}\no\\
&& + \int_{x,u,v}{\delta J^a(u) \over \delta A^c_i(x)}{\delta
J^b(v)\over \delta A^c_i(x)} ~{\d \over \d J^a( u) }{\d \over \d J^b(v) }\Psi
\Biggr]\no\\ 
V&&\!\!\!\!\!\!\!\!\!={1\over 2e^2} \int B^aB^a  \label{12}
\end{eqnarray}
where $B^a= \half \epsilon_{ij} (\partial_i A_j^a - \partial_j A_i^a +f^{abc}A_i^b
A_j^c)$.
Regularization is important in calculating the coefficients of the two terms in $T$.
Carrying this out we find
\begin{eqnarray}
T&&= m \Biggl[ \int_u J^a(u) {\d \over \d J^a(u)}\no\\
&&~~~~~ +\int \Omega_{ab} (u,v) 
{\d \over \d J^a(u) }{\d \over \d J^b(v) }\Biggr] \label{13a}\\
V&&={ \pi \over {m c_A}} \int \bdel J_a (\vx)
\bdel J_a (\vx) \label{13b}
\end{eqnarray}
where $m= e^2c_A/2\pi$ and
\be
\Omega_{ab}(u,v) = {c_A\over \pi^2} {\d_{ab} \over (u-v)^2} ~-~ 
i {f_{abc} J^c (v)\over {\pi (u-v)}}
\ee
The first term in $T$ shows that every power of $J$ in the wavefunction gives
a value $m$ to the energy, suggesting the existence of a mass gap.
The calculation of this term involves exactly the same quantity as in (\ref{7b})
and with the same regulator leads to (\ref{13a}), i.e.,
\begin{eqnarray}
-{e^2\over 2}&&\!\!\!\!\!\!\!\!\!\int d^2y {\delta^2 J_a(x)\over 
\delta \bA^b(y) \delta A^b(y)}\no\\
 &&={e^2c_A\over 2\pi} M^{\dag} _{am} \Tr \left[ T^m
\bD ^{-1}(y,x) 
\right]_{y \rightarrow x}\no\\
&&= m ~J_a(x)\label{13c}
\end{eqnarray}

Finally, (\ref{13a},\ref{13b}) (with regularizations taken account of) give a self-adjoint Hamiltonian
which, as I mentioned before, is a nice consistency check.

\section{THE VACUUM STATE}

Let us now consider the eigenstates of the theory.
The vacuum wavefunction is presumably the simplest to calculate.
Ignoring the potential term $V$ for the moment, since $T$ involves derivatives, 
we see immediately that
the ground state wavefunction for $T$ is $\Phi_0 =1$. This may seem like a 
trivial statement, but the key
point is that it is normalizable with the inner product (\ref{10}); 
in fact, the normalization integral is
just the partition function for the WZW action and is finite.
Starting with this, we can
solve the Schr\"odinger equation taking $\Psi_0$ to be of the form $\exp (P)$, where
$P$ is a perturbative series in the potential term $V$ (equivalent to a
$1/m$-expansion). We then get
\begin{eqnarray}
P =&&\!\!\!\!\!\!\!\!\! - {\pi \over { m^2 c_A}} \Tr \int  : \bdel J \bdel J : \no\\
&&\!\!\!\!\!\!\!\!\! - \left({\pi \over { m^2 c_A}}\right)^2 \Tr \int   \bigl[:
\bdel J ( {\cal D} \bd ) \bdel J 
    \no\\
&&\!\!\!\!+  {1 \over 3} \bdel J  [J, \bdel ^2 J] : \bigr] \no\\
&&\!\!\!\!\!\!\!\!\! - 2 \left({\pi \over {m^2 c_A}}\right)^3 \Tr \int \bigl[ :
\bdel J  ( {\cal D} \bd )^2 \bd J \no\\
&&\!\!\!\!\!\!\!\!\!+{2 \over 9} [ {\cal D} \bd J,~ \bd J] \bd ^2 J
+ {8 \over 9} [{\cal D} \bd ^2 J,~ J] \bd ^2 J \no\\
&&\!\!\!\!\!\!\!\!\!- {1 \over 6} [J, ~ \bd J] [\bd J,~ \bd ^2 J] - {2 \over 9} [J,
\bd J][J, \bd ^3 J]: \bigr]\no\\ 
&&+ {\cal O} ( {1 \over m^8})\label{P}
\end{eqnarray}
where ${\cal D}h= ({c_A / \pi}) \del h -[J,h]$.
The series is naturally grouped as terms with 2 $ J$'s, terms with 3 $J$'s, etc.
These terms can be summed up; for the $2J$-terms we find
\begin{eqnarray}
\Psi_0&&\!\!\!\!\!\!\!\!\!= \exp\left[ P\right]\no\\
P&&\!\!\!\!\!\!\!\!\!=-{1 \over {2 e^2}} \int_{x,y} B_a(x) K(x,y) B_a(y)
+ {\cal O}(3J)\no\\
K(x,y)&&\!\!\!\!\!\!\!\!\!=\left[{ 1 \over  {\bigl( m + 
\sqrt{m^2 - \nabla ^2 } \bigr)} }\right] _{x,y}\no\\
\label{wavfn}
\end{eqnarray}
The first term in (\ref{wavfn}) has the correct (perturbative) high momentum
limit, viz.,
\begin{eqnarray}
\Psi_0&&\!\!\!\!\!\!\!\!\!\approx  \exp\left[ -{1 \over {2 e^2}} \int_{x,y} B_a(x)
\Biggl[{ 1 \over
\sqrt{ - \nabla ^2 } }\right] _{x,y} B_a(y)\no\\
&& ~~~~~~~~~+ {\cal O}(3J)\Biggr]
\label{wavfn2}
\end{eqnarray}
Thus although we started with the high $m$ (or low momentum)
limit, the result (\ref{wavfn}) does match onto the perturbative limit.
The higher terms are also small for the low momentum limit.

We can now use this result to calculate the expectation value of the
Wilson loop operator which is given as
\begin{eqnarray}
W(C)&&
\!\!\!\!\!\!\!\!\!= \Tr P ~e^{-\oint_C (Adz+\bA d{\bar z})}\no\\
&&\!\!\!\!\!\!\!\!\!=
 \Tr P ~e^{(\pi /c_A)\oint_C J }
\label{11}
\end{eqnarray}
For the fundamental representation,
its expectation value is given by
\begin{eqnarray}
\la W_F (C) \ra &&= {\rm constant}~~\exp \left[ - \sigma {\cal
A}_C \right]\no\\
{\sqrt{\sigma }}&&= e^2 \sqrt{{N^2-1\over 8\pi}}
\label{tension}
\end{eqnarray}
where ${\cal A}_C$ is the area of the loop $C$. $\sigma$ is the string tension.
This is a prediction of our analysis starting from first principles with no adjustable
parameters. Notice that the dependence on $e^2$ and $N$ is
in agreement 
with large-$N$ expectations, with $\sigma$ depending only on the combination
$e^2N$ as $N\rightarrow \infty$. (The first correction to the large-$N$ limit
is negative, viz., $-(e^2N)/2N^2\sqrt{8\pi}$ which may be interesting in the context
of large-$N$ analyses.)
Formula (\ref{tension}) gives the values $\sqrt{\sigma}/e^2
=0.345, 0.564, 0.772, 0.977$ for $N=2,3,4,5$. 
There are estimates for $\sigma$ based on Monte Carlo simulations of lattice gauge
theory. The results for the gauge  groups $SU(2),~SU(3),
~SU(4)$ and $SU(5)$ are 
$\sqrt{\sigma}/e^2 =$ 0.335, 0.553, 0.758, 0.966 \cite{teper}. We see that
our result agrees with the lattice result to within $\sim 3\%$.

One might wonder at this stage why the result is so good when we have
not included the $3J$- and higher terms in the the wavefunction.
This is basically because the
string tension is determined by large area loops and for these, it is the
long  distance part of the wavefunction which contributes significantly.
In this limit, the $3J$- and higher terms in (\ref{wavfn}) are small compared to the
quadratic term. We expect their contribution to $\sigma$ to be small as well; this is
currently under study.

We have summed up the $3J$-terms as well. Generally, one finds that $P$,
when expressed in terms of the magnetic field, is nonlocal
even in a $(1/m)$-expansion, contrary
to what one might expect for a theory with a mass. This is essentially due to 
gauge invariance combined with our choice of $A_0=0$; it has recently been shown that
a similar result holds for the Schwinger model \cite{mansfield}.

\section{MAGNETIC MASS}

I shall now briefly return to the magnetic mass. From the expression (\ref{13a})
we see immediately that for a wavefunction which is just $J^a$, we have the exact
result $T ~J^a = m ~J^a$. 
When the potential term is added, $J^a$ is no longer an exact eigenstate; we find
\be
(T+V) ~J^a = \sqrt{m^2 -\nabla^2}~ J^a ~+~\cdots
\ee
showing how the mass value is corrected to the relativistic dispersion relation.

Now $J^a$ may be considered as the gauge-invariant definition
of the gluon. This result thus suggests a dynamical propagator
mass $m= e^2c_A/2\pi$ for the gluon.
A different way to see this result is as follows.
We can expand the matrix field $J$ in powers of $\vf_a$ which parametrizes $H$, 
so that
$J\simeq (c_A/\pi ) \partial \vf_a t_a$. This is like a perturbation expansion, but
a resummed or improved version of it, where we expand the WZW action in
$\exp(2c_A\S (H))$ but not expand the exponential itself. The Hamiltonian can then be
simplified as
\begin{eqnarray}
\H &&\!\!\!\!\!\!\!\!\!\simeq \half \int_x [- {\d ^2 \over {\d \phi _a ^2 (x)}} +
\phi_a (x)  \bigl( m^2 -
\nabla ^2 \bigr)  \phi_a (x)]\no\\
&&~~~~~~~~~~~ + ...\label{hamil2}
\end{eqnarray}
where $\phi_a (k) \!=\! \sqrt {{c_A k \bar{k} }/ (2 \pi m)} \vf _a (k)$, in momentum space.
In arriving at this expression we have expanded the currents and also absorbed the
WZW-action part of the measure into the definition of the wavefunctions, i.e.,
the operator (\ref{hamil2}) acts on 
${\tilde \Psi} =e^{c_AS(H)}\Psi$. 
The above equation shows that the propagating particles in the
perturbative regime, where the power series expansion of the current is
appropriate, have a mass $m=e^2c_A/2\pi$. 
This value can therefore be identified as the magnetic
mass of the gluons as given by this  nonperturbative analysis.

For $SU(2)$ our result is $m\approx 0.32 e^2$.  
Gauge-invariant resummations of perturbation theory have given the values
$0.38e^2$ \cite{AN} and $0.28e^2$ \cite{owe1}. 

Lattice estimates of this mass are
$0.31e^2$ to $0.40e^2$ (as a common factor mass for glueballs \cite{owe2}) and
$ 0.44e^2$ to
$0.46e^2$ in different gauges \cite{karsch}. 
An unambiguous lattice evaluation of this would require a gauge-invariant
definition a gluon operator and its correlator. It is worth pointing out that,
eventhough most lattice analyses only discuss glueballs, a gluon correlator is entirely
reasonable. After all, in perturbation theory, the pole or gap in the gluon propagator
is gauge-invariant to all orders and one might ask what the lattice version of this
quantity is. Of course, a gauge-invariant gluon operator cannot be defined as a 
local operator, this is evident from the definition of $J$. Philipsen has recently
given a nice definition of the ``gluon" on the lattice. Preliminary
numerical estimates
then give the mass as $0.37e^2$ \cite{owe3}.

\section{EXCITED STATES}

Eventhough $J$ is useful as  a description of the gluon, it is not a physical state.
This is because of an ambiguity in our parametrization (\ref{1}). Notice that
the matrices $M$ and $M{\bar V}(\bz )$ both give the same $A, \bA$, where
${\bar V}(\bz )$ only depends on $\bz$ and not $z$.  Since we have the same potentials,
physical results must be insensitive to this redundancy in the choice of $M$;
in other words, physical wavefunctions must be invariant under $M\rightarrow
M {\bar V}(\bz )$. $J$ is not invariant; we need at least two $J$'s to form
an invariant combination. An example is
\begin{eqnarray}
\Psi _2 &&\!\!\!\!\!\!\!\!\!= \int _{x,y} f(x,y) \times\no\\
&&\bigl[ \bdel J_a (x) \bigl(
H(x,\by) H^{-1} (y,
\by) \bigr) _{ab} \bdel J_b (y)  \bigr]\no\\
\end{eqnarray}
This is not quite an eigenstate of the Hamiltonian.
Neglecting certain ${\cal O}(J^3)$-terms, one can show that this is an approximate
eigenstate of eigenvalue $E$ if
\begin{eqnarray}
\Biggl[ &&\!\!\!\!\!\!\!\! \sqrt{m^2-\nabla_1^2}~ +\sqrt{m^2-\nabla_2^2} +2m\no\\
&&\!\!\!+\log (\vert x-y\vert^2/\lambda )\Biggr] f(x,y)= E ~f(x,y)\no\\
\label{glueball}
\end{eqnarray}
An {\it a posteriori} justification of this would require that the size of the bound
state be not too large on the scale of $1/m$. This does not seem to be the case, so, at
least for now, all we can say is that the mass of $\Psi_2 \geq 2m$; 
see  however \cite{2}.

\section{YANG-MILLS-CHERN-SIMONS THEORY}

We have extended our analysis to the Yang-Mills-Chern-Simons theory
by adding a level $k$ Chern-Simons term to the action, 
which gives a perturbative mass
$e^2/4\pi$ to the gluon \cite{YMCS}. The inner product now becomes
\be
\la 1 | 2\ra = \int d\mu (H) e^{(k+2c_A)~\S (H)}~~\Psi_1^* \Psi_2 
\label{ymcs1}
\ee
The earlier conformal field theory argument shows that there are now new integrable
operators. These lead to screening behaviour for the Wilson loop operator
rather than confinement, as expected. The kinetic energy term becomes
\begin{eqnarray}
T&&= {e^2\over 4\pi}(k+2c_A) \int_u J^a(u) {\d \over \d J^a(u)}\no\\
&&~~~~~ +{e^2c_A\over 2\pi}\int \Omega_{ab} (u,v) 
{\d \over \d J^a(u) }{\d \over \d J^b(v) }\no\\
\label{ymcs2}
\end{eqnarray}
The gluon mass is now $(k+2c_A)e^2/4\pi$ reflecting dynamical 
mass generation as well
by nonperturbative effects. The vacuum state and a number of excited states
have also been constructed in this case \cite{YMCS}.


\begin{thebibliography}{99}

\bibitem{1}
D. Karabali and V.P. Nair, Nucl. Phys. B464 (1996) 135; Phys. Lett.
B379 (1996) 141; Int. J. Mod. Phys. A12 (1997) 1161.
\bibitem{2}
D. Karabali, Chanju Kim and V.P. Nair, Nucl. Phys. B524 (1998) 661.
\bibitem{3} 
D. Karabali, Chanju Kim and V.P. Nair, Phys. Lett. B434 (1998) 103.
\bibitem{lingpy}
A.D. Linde, Phys. Lett. B96 (1980) 289;
D. Gross, R. Pisarski and L. Yaffe, Rev. Mod. Phys. 53 (1981) 43.
\bibitem{poly}
A.M. Polyakov and P.B. Wiegmann, Phys.Lett. B141 (1984) 223.
\bibitem{witt}
E. Witten, Commun. Math. Phys. 92 (1984) 455;
S.P. Novikov, Usp. Mat. Nauk. 37 (1982) 3.
\bibitem{GKBN}
K. Gawedzki and A. Kupiainen, Phys. Lett. B215 (1988) 119;
Nucl.Phys. B320 (1989) 649; M. Bos and V.P. Nair, 
Int.J.Mod.Phys. A5 (1990) 959.
\bibitem{teper}
M. Teper, Phys. Lett. B311 (1993) 223; O. Philipsen,
M. Teper and H. Wittig, Nucl.Phys. B469 (1996) 445;
M. Teper, hep-lat/9804008 and references therein.
\bibitem{mansfield}
D. Nolland and P. Mansfield, Int. J. Mod. Phys. A15 (2000) 429.
\bibitem{AN}
G. Alexanian and V.P. Nair,
Phys.Lett. B352 (1995) 435;
\bibitem{owe1}W. Buchm\"uller and O. Philipsen, Nucl.Phys. B443 (1995) 47;
R. Jackiw and S.Y. Pi, Phys.Lett. B368 (1996) 131;
{\it ibid} B403 (1997) 297. 
\bibitem{owe2}
W. Buchm\"uller and O. Philipsen, Phys. Lett. B397 (1997) 112.
\bibitem{karsch}
F. Karsch {\it et al}, Nucl. Phys. B474 (1996) 217;
F. Karsch, M. Oevers and P. Petreczky, Phys. Lett. B442 (1998) 291.
\bibitem{owe3}
O. Philipsen, Phys. Lett. B521 (2001) 273; talk at {\it Lattice 2001},
hep-lat/0110114.
\bibitem{YMCS}
D. Karabali, Chanju Kim and V.P. Nair, Nucl. Phys. B566 (2000) 331.
 
\end{thebibliography}
\end{document}